\documentclass[conference]{IEEEtran}
\IEEEoverridecommandlockouts
\usepackage{color,soul}
\usepackage{cancel}
\usepackage{multirow}
\usepackage{cite}
\usepackage{amsmath,amssymb,amsfonts, bm}
\usepackage{algorithmic}
\usepackage{graphicx} 
\usepackage{textcomp}
\usepackage{xcolor}
\def\BibTeX{{\rm B\kern-.05em{\sc i\kern-.025em b}\kern-.08em
    T\kern-.1667em\lower.7ex\hbox{E}\kern-.125emX}}

\newcommand*{\vertbar}{\rule[-1ex]{0.5pt}{2.5ex}}

\begin{document}

\title{A Learning-Based Approach for Bias Elimination in Low-Cost Gyroscopes\\
}

\author{\IEEEauthorblockN{Daniel Engelsman}
\IEEEauthorblockA{\textit{The Hatter Department of Marine Technologies} \\
\textit{University of Haifa}\\
Haifa, Israel \\
dengelsm@campus.haifa.ac.il
}
\and
\IEEEauthorblockN{Itzik Klein, Senior Member, IEEE}
\IEEEauthorblockA{\textit{The Hatter Department of Marine Technologies} \\
\textit{University of Haifa}\\
Haifa, Israel \\
kitzik@univ.haifa.ac.il
}
}

\maketitle

\begin{abstract}
Modern sensors play a pivotal role in many operating platforms, as they manage to track the platform dynamics at a relatively low manufacturing costs. Their widespread use can be found starting from autonomous vehicles, through tactical platforms, and ending with household appliances in daily use. Upon leaving the factory, the calibrated sensor starts accumulating different error sources which slowly wear out its precision and reliability. To that end, periodic calibration is needed, to restore intrinsic parameters and realign its readings with the ground truth. While extensive analytic methods exist in the literature, little is proposed using data-driven techniques and their unprecedented approximation capabilities. In this study, we show how bias elimination in low-cost gyroscopes can be performed in considerably shorter operative time, using a unique convolutional neural network structure. The strict constraints of traditional methods are replaced by a learning-based regression which spares the time-consuming averaging time, exhibiting efficient sifting of background noise from the actual bias.
\end{abstract}

\begin{IEEEkeywords}
Inertial Measurement Units, Gyroscopes, Calibration, State Estimation, Deep Learning
\end{IEEEkeywords}

\section{Introduction}
Inertial sensors are subjected to a wide range of measurement errors which obscure the meaningful signal and degrade their reliability. Uncalibrated sensors are characterized by outputting wrongful measurements, a phenomenon that worsens when these quantities are integrated, to obtain the navigation solution. These deviations can be corrected by low-pass filters or state estimation algorithms (e.g. Kalman filters), however knowledge of initial errors is required apriori, to prevent its projection on other state variables. Noise reduction can be obtained by averaging measurements from multiple IMUs, and similarly, bias can be mitigated by subtracting opposite polarities of an IMU pair, mounted on the same axis. However, often this redundancy is not achievable or just not affordable, thus calibration is inevitable, especially when platforms are exclusively dependent on the sensor performance. 

Calibration refers to the task of determining the relationship between measurand outputs and their corresponding references, which serve as ground-truth (GT) \cite{b_barbour}. Following the calculation procedure, the sensor intrinsic parameters can be readjusted to minimize unwanted error sources, thus improving its performance. Model-based approaches offer an in-depth analysis of the error origins using different statistics and signal processing tools \cite{b_bekkeng, b_aggarwal}. Given a new measurement, the errors are identified and compensated, depending on how well they align with the model. In ideal noise-free conditions, a single time step is all that is required to determine the null bias, by simply differencing the output from zero. However realistically, minimum error is largely affected by the averaging time period, as noise suppression exhibits a slow error decay. This approach faces another obstacle when sterile conditions cannot be met, as in real-time scenarios, when platform is subjected to disruptions, giving rise to misestimated parameters. Recently, data-driven approaches showed superior performance over model-based methods in several navigation related tasks \cite{b_klein}. For example, an end-to-end deep-learning framework was shown to improve velocity estimation of an autonomous underwater vehicle \cite{b_cohen}. In pedestrian dead reckoning, learning approaches outperform model-based approaches, offering better position accuracy \cite{b_asraf, b_trigoni}, and mode classification performance \cite{b_shavit}. Recently, data-driven method was driven for denoising of accelerometer readings \cite{b_engelsman}. 

Therefore, to solve the calibration problem, this paper introduces a data-driven method, capable of estimating GT biases using supervised learning approach. By doing so, we show that: (i) non-linear estimators, i.e. convolutional neural networks (CNN), can outperform linear model-based calibration (ii) the well-trained model can reduce averaging times by one order of magnitude (iii) learnable parameters enhance model robustness to non-stationary disturbances.

The rest of the paper is organized as follows: Section \ref{s:gyro_model} elaborates on the gyroscope error model, Section \ref{s:methodology} describes our strategy and the proposed solution, Section \ref{s:experiment} presents analysis and results and Section \ref{s:conclusions} gives conclusions.

\section{Gyroscope Error Model} \label{s:gyro_model}
Sensor errors are generally classified into two categories: stochastic and deterministic sources. The first type is unmodelable due to unknown origin, changes randomly between time steps but overall variance remains constant when remeasured. The second type is modelable, constant between time steps but may vary when remeasured \cite{b_woodman}. The linear relationship between the gyro outputs $\boldsymbol{\tilde{\omega}}^b_{ib}$ and the true angular rates $\boldsymbol{\omega}^b_{ib}$, contains both errors, and is commonly modeled as
\begin{align} \label{eq:system}
\boldsymbol{\tilde{\omega}}^b_{ib} = \boldsymbol{M} \boldsymbol{\omega}^b_{ib} + \textbf{\textit{b}}_g + \textit{\textbf{w}}_g
\end{align}
where $\textbf{\textit{b}}_g$ is a constant bias, $\textit{\textbf{w}}_g$ is an i.i.d zero-mean Gaussian noise and $\boldsymbol{M}$ is a combination of the scale-factor (SF) diagonal matrix and the misalignment (MA) skew symmetric matrix
\begin{align}
\boldsymbol{M} = (\boldsymbol{I}_3 + \bm{M}_{\textbf{SF}} + \bm{M}_{\textbf{MA}})
\end{align}

Identification of the stochastic errors is commonly performed via Allan Variance analysis in the time domain \cite{b_av}, or power spectral density (PSD) in the frequency domain \cite{b_psd}. In contrast, deterministic errors are represented by a linear mapping of the input references to their corresponding outputs. Knowledge of their inherent nature is required for the calibration model, such that errors from raw measurements could be identified and compensated in accordance. 

\subsection{Stochastic Error Sources} 
Sensor noise originated in different random processes, are characterized by different PSDs. According to IEEE-STD-952 \cite{b_IEEE}, the five major error sources present in inertial sensors are quantization noise Q, angle/velocity random walk N, bias instability or $1/f$ noise B, rate random walk K and rate ramp R. Given no statistical dependence between each of the random processes, the total sum of their variances can be viewed as
\begin{align} \label{eq:noise}
\sigma^2_{\text{tot.}}  = \sigma^2_{\text{Q}} + \sigma^2_{\text{N}} + \sigma^2_{\text{B}} + \sigma^2_{\text{K}} + \sigma^2_{\text{R}} 
\end{align}

The first two terms are commonly fused together into a white Gaussian noise by normalizing with the square root of the time-step $\sqrt{\Delta t}$, and the last two terms are negligible in short averaging times, as in this study. However the bias instability $\sigma_{B}$, turns out to be the most problematic factor for the stationarity assumptions, as it indicates the maximum operation period before statistical properties (mostly mean and variance) cease to be time-invariant, impairing the applicability of estimation filters. Avoiding this drift is possible by simply limiting durations to one minute \cite{b_Groves}, where dynamic component is approximated to be less than tenth of the gyro noise 
\begin{align}
{\textbf{\textit{b}}}_g \triangleq {\textbf{\textit{b}}}_{g,s} + \cancelto{0}{ {\textbf{\textit{b}}}_{g,d} }
\end{align} 

To summarize, stochastic error sources in this study do not affect the bias term. Improving signal-to-noise ratio (SNR) is enabled by long averaging times, as overall noise magnitude is subjected only to the mean of first two noise sources in \eqref{eq:noise}
\begin{align}
\bar{\textit{\textbf{w}}}_g (t) = \operatorname{E} \big[ \textit{\textbf{w}}_g (t) \big] \quad , \quad \textit{\textbf{w}}_g \sim \sum_i \mathcal{N} (0, \, \sigma_i^{2} )
\end{align}

\subsection{Deterministic Error Sources}
As already mentioned in \eqref{eq:system}, true angular rates are scaled by the gain (slope) matrix $\boldsymbol{M}$, but their offset from the origin (intercept) is governed by true fixed biases $\textbf{\textit{b}}_g$, disturbed by background noise $\bar{\textit{\textbf{w}}}_g $. Under steady-state conditions, the system in matrix form is given by
\begin{align}
\begin{bmatrix} \label{eq:system_2}
\tilde{\omega}_{x} \\ \tilde{\omega}_{y} \\ \tilde{\omega}_{z}
\end{bmatrix} = \left[ \begin{array}{ccc}
m_{xx} & m_{xy} & m_{xz}\\
m_{yx} & m_{yy} & m_{yz}\\
m_{zx} & m_{zy} & m_{zz}
\end{array} \right] \begin{bmatrix}
\omega_{x} \\ \omega_{y} \\ \omega_{z}
\end{bmatrix} + \begin{bmatrix} 
b_x \\ b_y \\ b_z
\end{bmatrix} + 
\begin{bmatrix}
\bar{\textit{w}}_x (t) \\ \bar{\textit{w}}_y (t) \\ \bar{\textit{w}}_z (t)
\end{bmatrix}
\end{align}

Since the last error is a function of time, it can be conveniently added to the general bias term as follows, yielding
\begin{align} \label{eq:bias}
\bar{\textbf{\textit{b}}}_g (t) &= \textbf{\textit{b}}_g + \bar{\textit{\textbf{w}}}_g (t)
\end{align}
implying that bias precision requires low bias error (residual), and is also determined by the averaging time period
\begin{align} \label{eq:residual}
\boldsymbol{\delta} {\textbf{\textit{b}}}_g (t) &= \bar{\textbf{\textit{b}}}_g (t) - {\textbf{\textit{b}}}_g = \bar{\textit{\textbf{w}}}_g (t)
\end{align}
such that the linear system, \eqref{eq:system_2}, can be augmented into
\begin{align}
\boldsymbol{\tilde{\omega}}^b_{ib} = [\boldsymbol{M} \, | \, \bar{\textbf{\textit{b}}}_g ] \begin{bmatrix} \boldsymbol{{\omega}}^b_{ib} \\ 1 \end{bmatrix}
\end{align}
Next, finding the twelve calibration parameters is enabled by rearranging the equations into a linear system $\textbf{Ax}=\textbf{b}$ whose coefficient matrix is $\textbf{A} \in \mathbb{R}^{3 \times 12}$ and the calibration vector is given by $\textbf{x} \in \mathbb{R}^{12}$. A single measurement, with known input and measurable output, provides the following system \cite{b_calibration}

\begin{align} \label{eq:vec}
\begin{bmatrix}
\vertbar \\  \boldsymbol{\tilde{\omega}}^b_{ib} \\ \vertbar \end{bmatrix} = 
\left[ \begin{array} {ccc|c}
(\boldsymbol{{\omega}}^b_{ib})^\top & \boldsymbol{0}_{1 \times 3} & \boldsymbol{0}_{1 \times 3} & \\
\boldsymbol{0}_{1 \times 3} & (\boldsymbol{\omega}^b_{ib})^\top & \boldsymbol{0}_{1 \times 3} &  \boldsymbol{I}_{3 \times 3} \\
\boldsymbol{0}_{1 \times 3} & \boldsymbol{0}_{1 \times 3} & (\boldsymbol{\omega}^b_{ib})^\top & 
\end{array} \right]
\begin{bmatrix}
\vertbar \\ \boldsymbol{m} \\ \vertbar \\[1.5mm] \bar{\textbf{\textit{b}}}_g \end{bmatrix}
\end{align}
Using linear least squares, solution requires a minimum of four non-zero points $\boldsymbol{\omega}^b_{ib} \neq \textbf{0}$ to avoid a trivial solution of $\boldsymbol{m}=\textbf{0}$. However practically, solution can be obtained by collecting six different data points, using a precision turntable to provide known turning rates for each sensitive axis, in both clockwise and counter clockwise directions \cite{b_titterton}. This way the error resolution is dependent on the equipment ability to maintain steady state conditions $\frac{ \partial \boldsymbol{\omega} }{ \partial t} = \textbf{0}$, as non-stationary disturbances may be accumulated over the elapsed time. 

In contrast to $\boldsymbol{m}$, bias terms can be determined more easily, by a single stationary measurement of  $\boldsymbol{\omega}^b_{ib} = \textbf{0}$, since most low-cost gyros are insensitive to Earth rotation rate.

\section{Methodology} \label{s:methodology}
In supervised learning, model $f$ is required to map input samples $\textbf{\textit{x}}_i$ into their corresponding labels $\textbf{\textit{y}}_i$, while being iteratively assessed for its performances. This section elaborates on the steps taken from data preparation to model evaluation. 

\begin{figure}[h] 
\begin{center}
\includegraphics[width=0.5\textwidth]{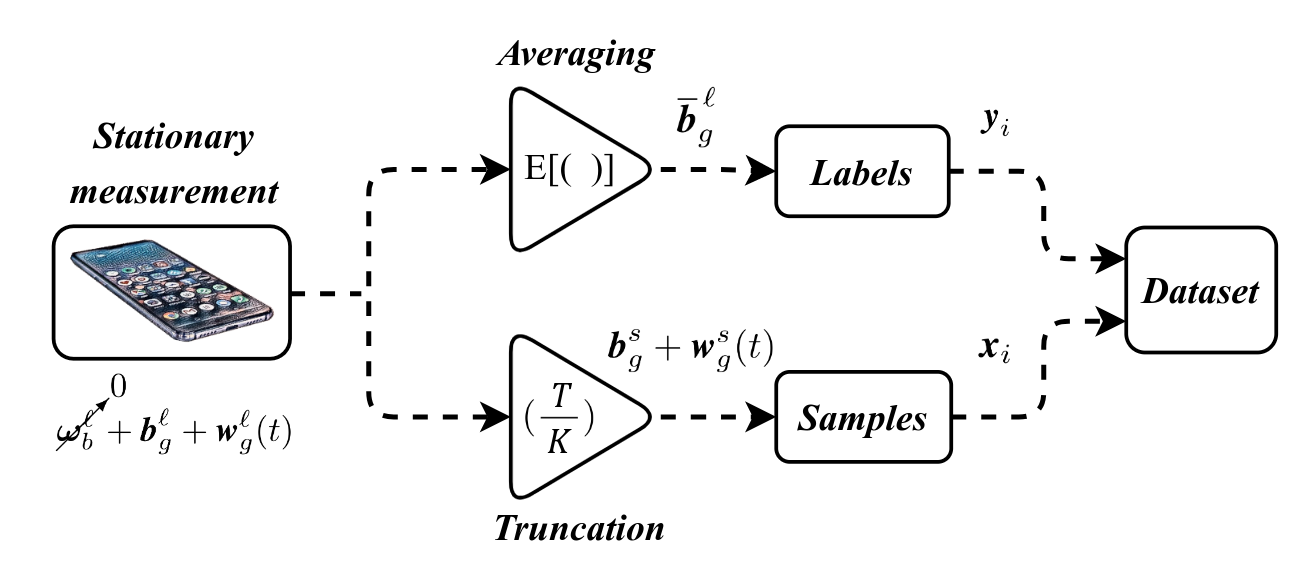}
\caption{Conceptual flow of the dataset generation.}
\label{f:system}
\end{center}
\end{figure}

\subsection{Data Acquisition}
Using the freely available phyphox app \cite{b_app}, the dataset was constructed of ten one-minute long ($\ell$) measurements, taken from five different low-cost inertial sensors sampled at 200$\,$Hz\cite{b_dataset}. To allow a wide range of fixed biases, different standby times taken before measurements, were commenced, to guarantee an in-run (repeatability) bias. Fig.~\ref{f:system} illustrates how raw inputs are converted into sample-label pair, consequently constructing a dataset whose train-test is divided by a split ratio of 80:20. The truncation operator divides total durations ($T$) by a division factor ($K$), to obtain shorter samples ($s$). The averaging operator returns the expected value of long measurements, to simulate the biases as accurately as possible, and then use them as corresponding labels.

\subsection{Augmentation}
To increase variability and reduce the risk of overfitting, each measurement has been copied a hundred times, then modified by two transformations suitable for time-series (order does not matter): additive bias and additive zero mean white noise, both are normally distributed. This way, the augmented dataset reached a total of 5,000 samples, improving the model generalizability and robustness to noise \cite{b_augment}.

\subsection{Neural Network Structure}
\begin{figure}[h]
\begin{center}
\includegraphics[width=0.46\textwidth]{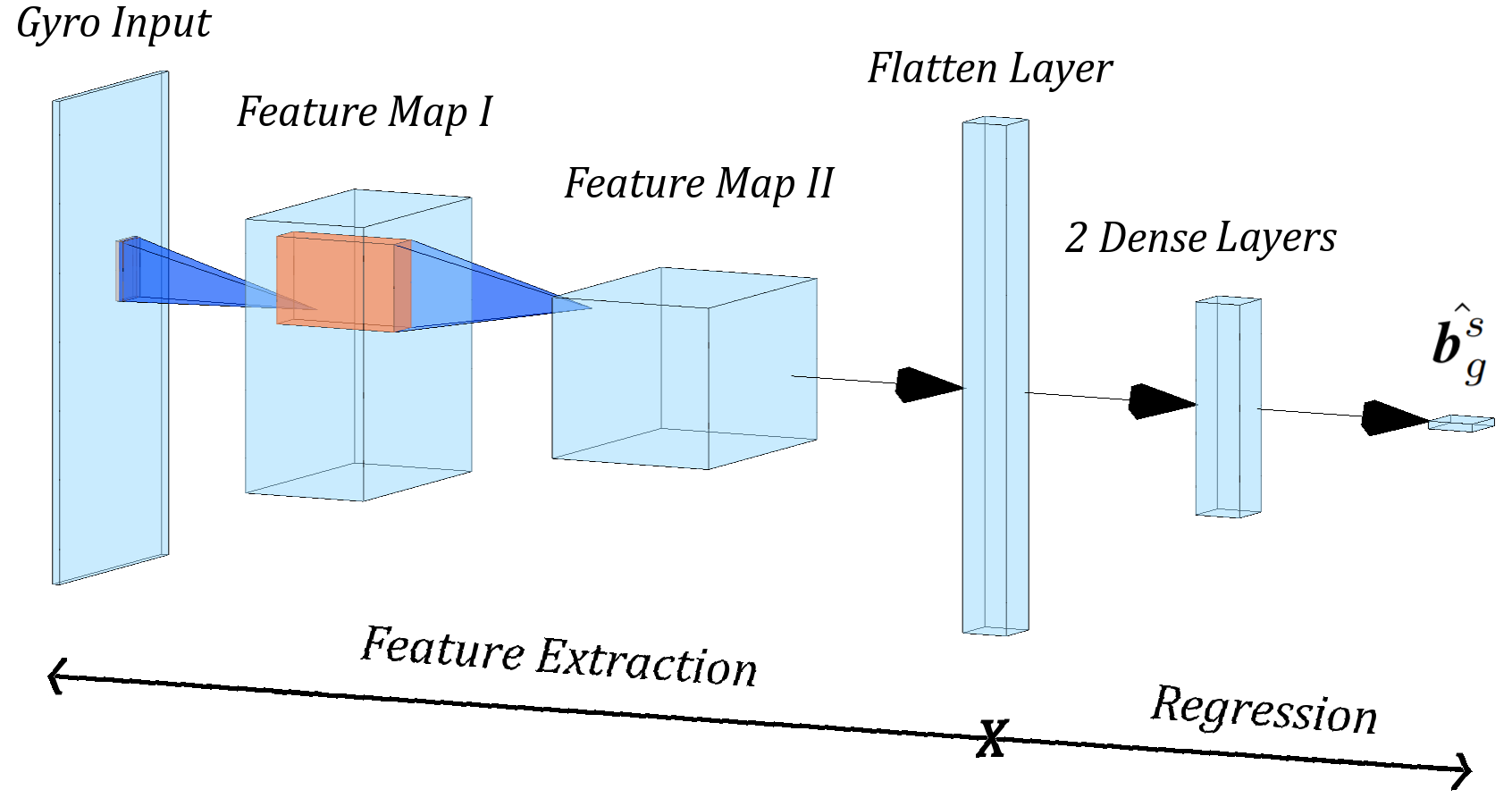}
\caption{Architecture of the proposed CNN model.}
\label{f:CNN}
\end{center}
\end{figure}

Since all input samples are equidimensional, and no complex dynamics are introduced, two-layer convolutional neural network (CNN) was chosen to operate as the approximation function. Fig.~\ref{f:CNN} illustrates its architecture, where encoder module performs sequential 1D convolutions, followed by subsampling (pooling) operation and ending with a low-dimensional features representation (feature map II). Then, regression phase is performed by a fully-connected network, which generates accurate bias predictions, denoted as $\, \hat{\textbf{\textit{b}}^s_g}$.

\subsection{Evaluation metric}
To assess the model's performance, deviations between the approximated output $\hat{\textbf{\textit{b}}^s_g } := f( \textbf{\textit{b}}^s_g )$ and their corresponding GT labels $\bar{\textbf{\textit{b}}}_g^{\ell}$, are given by the following error function
\begin{align}
\boldsymbol{e} = \hat{\textbf{\textit{b}}^s_{g} } -  \bar{\textbf{\textit{b}}}_{g}^{\ell} \ \in \ \mathbb{R}^{3}
\end{align}
During training phase the model learns by minimizing the above error. To that end, a common loss function is the mean squared error (MSE), allowing progressive tuning of vector weights $\boldsymbol{w}$, as it is continuously differentiable
\begin{align}
\boldsymbol{w}^* = \arg \min_{\boldsymbol{w}} \mathcal{L}_{\text{MSE}}
\quad , \quad 
\mathcal{L}_{\text{MSE}} = \operatorname{E} \{ \boldsymbol{e}^{\text{T}} \boldsymbol{e} \}
\end{align}

\section{Experiments and Results} \label{s:experiment}
To understand the model's contribution, the rolling mean which is used for comparison, will be explained using one of the measurement devices in the dataset, Samsung Galaxy S7. To allow for careful examination of the deviations, units in y-axes are given in milliradians and degrees. 
\begin{figure}[h]
\begin{center}
\includegraphics[width=0.475\textwidth]{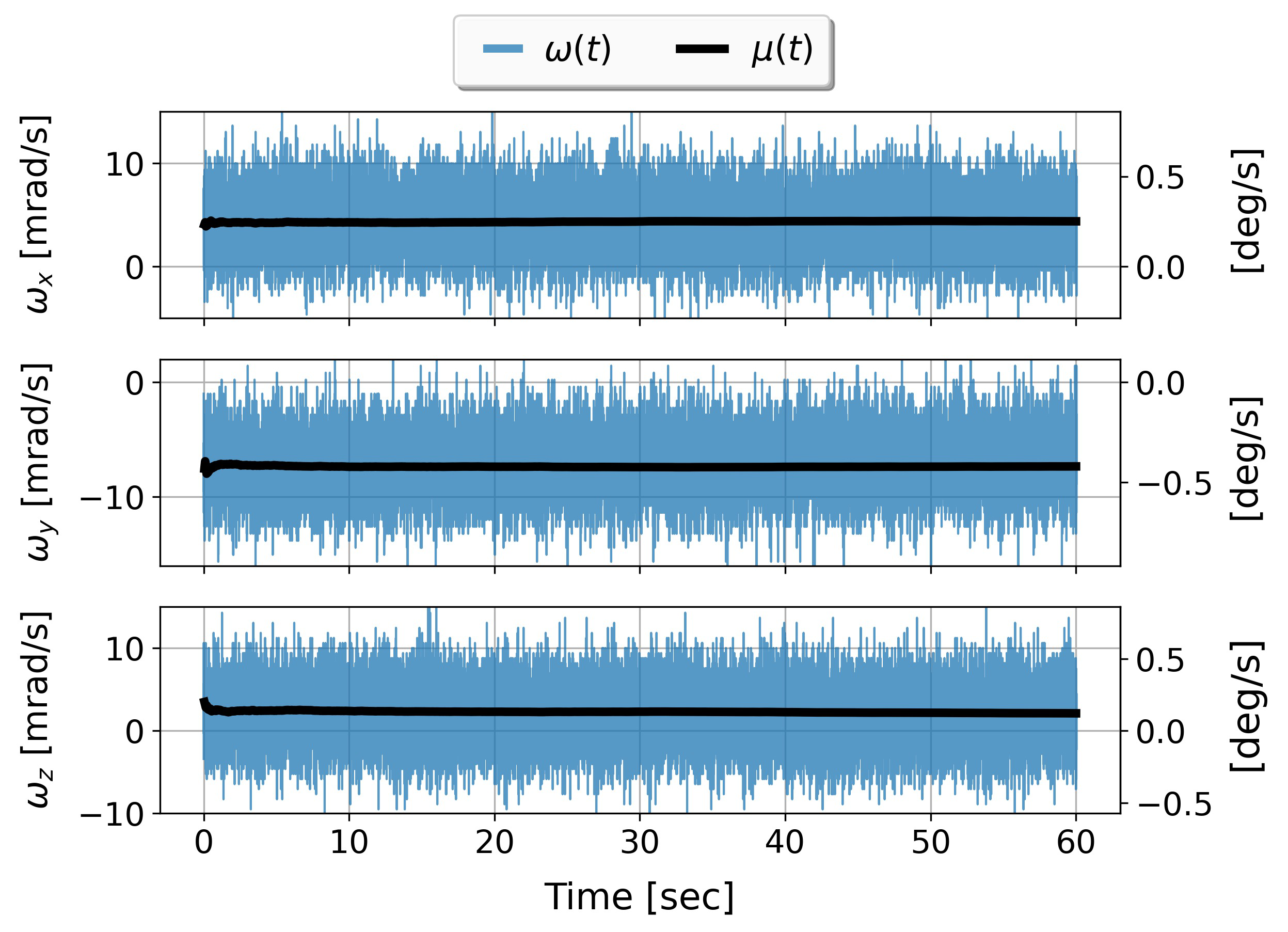}
\caption{Gyros outputs and their corresponding running mean.}
\label{f:gyros}
\end{center}
\end{figure}

Fig.~\ref{f:gyros} shows a one-minute measurement, where gyro outputs are given in bright blue and their calculated rolling means are in black, correlating with \eqref{eq:system} and \eqref{eq:bias}, respectively. In the absence of any clear trend, the center exhibits a slight offset from zero, while fluctuating within steady boundaries. 
\begin{figure}[h] 
\begin{center}
\includegraphics[width=0.475\textwidth]{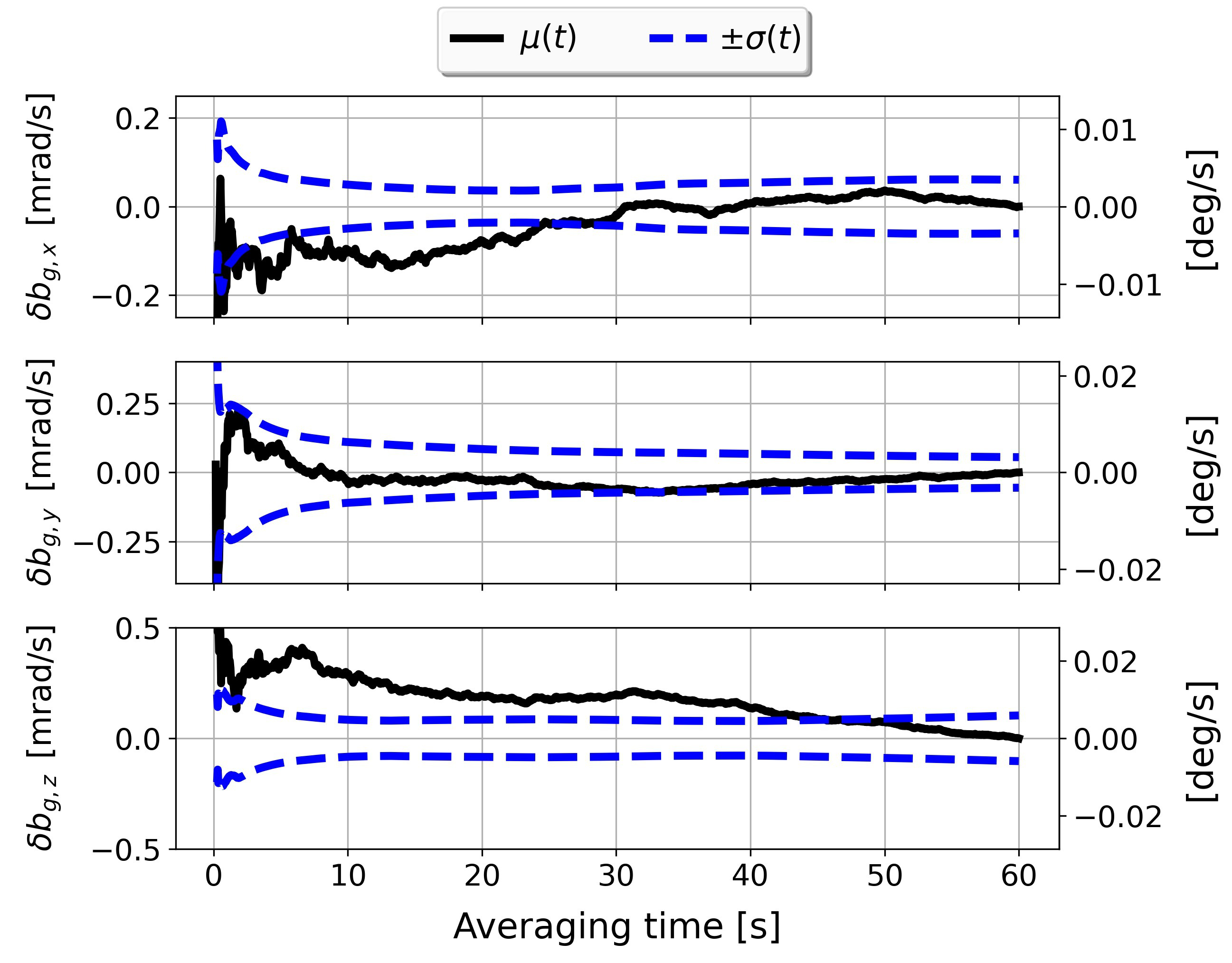}
\caption{Gyros residuals vs. averaging time.}
\label{f:residuals}
\end{center}
\end{figure}

Fig.~\ref{f:residuals} shows the dynamic decay of the bias error (residual) in black, and its sensitivity to the noise density, as stated in \eqref{eq:residual}. At first, errors exhibit an unstable behaviour, but as soon as averaging time approaches one minute, they are significantly reduced. The dashed blue margins denote the moving standard deviations, emphasizing the scattering dynamics of the distribution, and its attenuation with respect to elapsed time. Due to this time-consuming process, our learning-based framework proposes a competitive alternative, which approximates actual biases at notably shorter time intervals.

Next, to get a good impression of how the model outperforms simple averaging, i.e. baseline, the one minute scenario is divided by four division factors, $K$, creating four datasets. 
\begin{figure}[h] 
\begin{center} 
\includegraphics[width=0.5\textwidth]{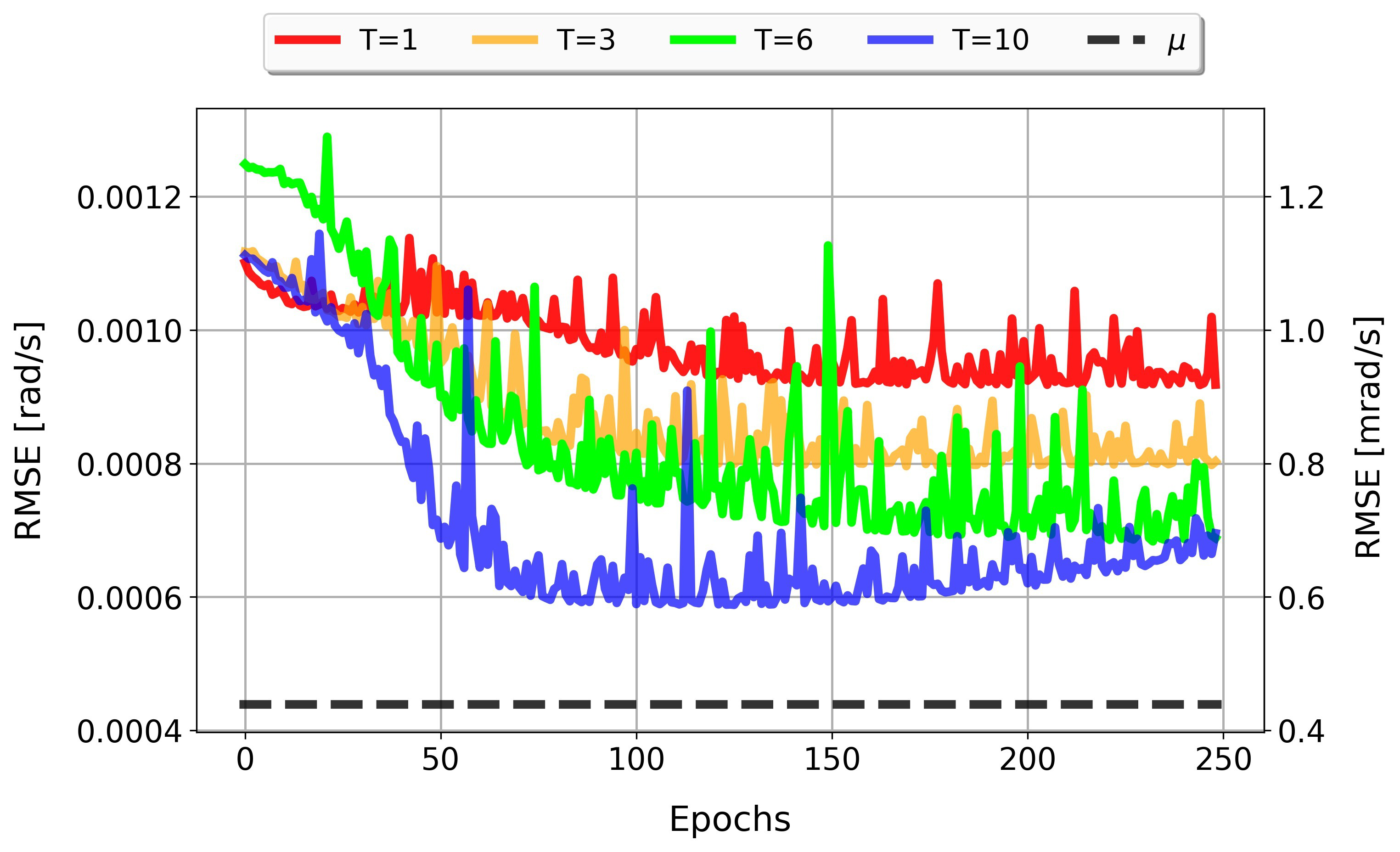}
\caption{Loss curves during training.} \label{f:validation}
\end{center}
\end{figure}

\begin{figure}[b]
\begin{center}
\includegraphics[width=0.5\textwidth]{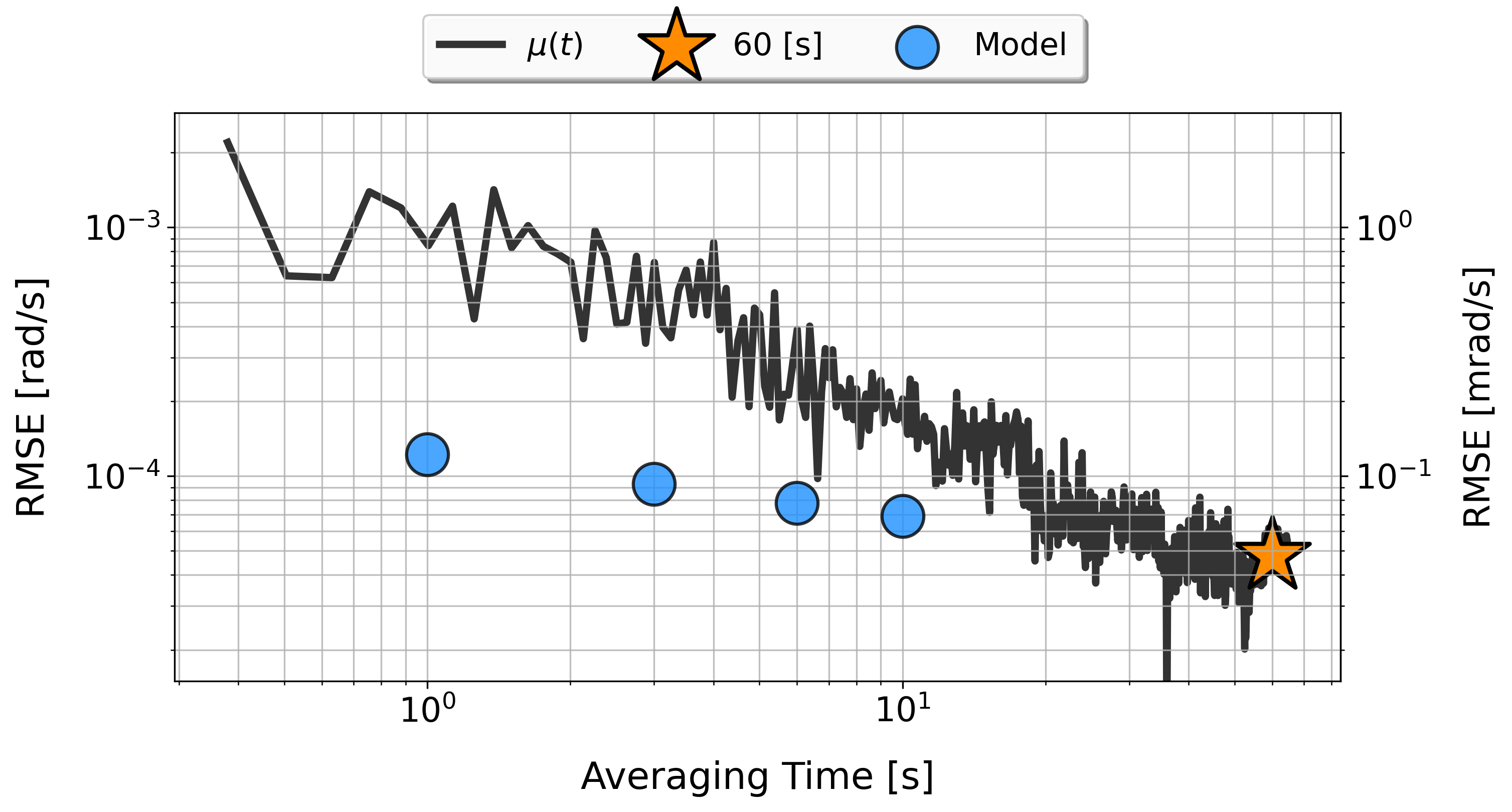}
\caption{Bias errors vs. averaging time.}
\label{f:model}
\end{center}
\end{figure}

Fig.~\ref{f:validation} presents all four training curves, in terms of square root of the loss function (RMSE), and the baseline is given at the bottom, by the black dashed line. The inverse relation between duration and error can be clearly seen, as longer durations improve the model approximations (due to lower loss). In Fig.~\ref{f:model}, the model capabilities of time reduction are demonstrated using an unseen sample of the Samsung S7. For comparison, model estimates are denoted by blue bubbles and the baseline by the black noisy curve. 

The orange star denotes the finite error after 60 seconds averaging, and naturally, loss refers to scalar summation of all three gyro triad. As seen in this example, but may vary given different noise regime, all model approximations are found below the baseline averaging when both are compared over equal-time steps. The bias residual decays slowly and exhibits sharp fluctuations whenever sudden noise bursts emerge. In contrast, learning-based models are able to "skip" these precious waiting times, and predict actual biases, as learned during their training phase. 

Finally, to assess a comprehensive error estimation over a test-set of 1,250 unseen samples, the following ratio is used
\begin{align} \label{eq:gamma}
\gamma := \left( \frac{ \varepsilon_{M} }{ \varepsilon_{B} } \right) \cdot 100 \ \text{[$\%$]} \quad , \quad \varepsilon = \text{RMSE}
\end{align}
and optimal models with minimum loss (Fig.~\ref{f:validation}), are taken for inference. This way, the relative contribution of each model $\varepsilon_M$, is compared with respect to the baseline $\varepsilon_{B}$.
\begin{table}[!h]
\renewcommand{\arraystretch}{1.25}
\caption{Bias RMSE vs. sample duration}
\begin{center} \label{t:results}
\begin{tabular}{c|c|c|c|c|} \cline{2-5} 
 & $K$ [-] & $T$ [sec] & RMSE [mrad/s] & $\gamma$ [$\%$] \\ \hline
\multicolumn{1}{|c|}{ \multirow{4}{*}{Models} } & 60 & 1.0 & 0.91759  & 208.73 \\ \cline{2-5}
\multicolumn{1}{ |c| }{}& 20 & 3.0 & 0.79598 & 181.07 \\ \cline{2-5}
\multicolumn{1}{ |c| }{}& 10 & 6.0 & 0.68328 & 155.43 \\ \cline{2-5}
\multicolumn{1}{ |c| }{}& 6 & 10.0 &  0.58821 & 133.80 \\ \hline
\multicolumn{1}{ |c| }{Baseline}   &  1 & 60.0 & 0.43959 & 100.00\\ \hline
\end{tabular}
\end{center}
\end{table}

Results in Table~\ref{t:results} corroborate the findings in Fig.~\ref{f:model}, 
suggesting a tradeoff between length of measurements and $\gamma$ coefficients, as longer durations achieve lower errors. However also here, moderate noise suppression seem to challenge the improvement rate. For example, results of one-second test-set is about double the baseline error. But when durations are extended by one order of magnitude (to 10 seconds), although $\gamma$ is eventually reduced, improvement rate is not proportional, but still one third larger than the baseline.

\section{Conclusions} \label{s:conclusions}
This paper addresses the challenging task of eliminating bias errors in low-cost gyros, often prone to wide range of noise. Despite stationary conditions and relatively minor biases, after one integration, angular errors are potential to grow linearly with time. Conventionally, noise averaging is a time-consuming process, used to extract their magnitude, and compensate them from corresponding state variables. However here, we demonstrated a time-sparing alternative, data-driven, using two-layer convolutional neural networks. Although sufficient data acquisition and model trainings are required, the advantages shown are clear. Biases can be determined significantly faster at the expense of a relatively small error, thus removing the need for precious averaging times at completely sterile environment.

\end{document}